\documentclass[apl,reprint,floatfix,a4paper]{revtex4-1}

\usepackage[T1]{fontenc}
\usepackage{etex}
\usepackage{geometry}
\usepackage{color,tabularx,colortbl}

\usepackage{setspace}
\usepackage{pgfplots}
\usepackage{booktabs}
\usepackage{tikz, tikz-3dplot}
\tikzset{>=latex}
\usepackage[percent]{overpic}
\pgfplotsset{compat=1.13}
\usepgfplotslibrary{external} 
\usepgfplotslibrary{fillbetween}
\usepgfplotslibrary{colorbrewer}
\usetikzlibrary{pgfplots.groupplots,arrows,decorations.markings}
\usetikzlibrary{shapes.geometric,calc,patterns}
\usepackage[strict]{changepage}

\usepackage{sidecap}
\sidecaptionvpos{figure}{t}

\usepackage{todonotes}
\makeatletter
\renewcommand{\todo}[2][]{\tikzexternaldisable\@todo[#1]{#2}\tikzexternalenable}
\renewcommand{\missingfigure}[2][]{\tikzexternaldisable\@missingfigure[#1]{#2}\tikzexternalenable}
\makeatother

\usepackage{dcolumn}
\newcolumntype{d}[1]{D{.}{.}{#1}}

\pgfmathsetmacro{\sigmaGaussFWHMlarge}{1.69864360058}
\pgfmathsetmacro{\sigmaGaussFWHMsmall}{0.849321800288}
\pgfmathsetmacro{\Tmin}{270}
\pgfmathsetmacro{\TClarge}{644.326893845}
\pgfmathsetmacro{\TCsmall}{536.939078204}

\definecolor{android_blue}{RGB}{51,181,229}
\definecolor{android_dark_blue}{RGB}{0,153,204}
\definecolor{android_pink}{RGB}{170,102,204}
\definecolor{android_purple}{RGB}{156,39,176}
\definecolor{android_dark_pink}{RGB}{153,51,204}
\definecolor{android_green}{RGB}{153,204,0}
\definecolor{android_dark_green}{RGB}{102,153,0}
\definecolor{android_orange}{RGB}{255,152,0}
\definecolor{android_dark_orange}{RGB}{255,152,0}
\definecolor{android_red}{RGB}{255,68,68}
\definecolor{android_dark_red}{RGB}{204,0,0}
\definecolor{android_pink}{RGB}{156,39,176}
\definecolor{android_grey}{RGB}{158,158,158}

\pgfplotsset{grid style={dashed,grey,opacity=0.5}}
	  
\pgfplotscreateplotcyclelist{peak_temp}{
	  {color=android_green,line width=0.5pt,mark size=2pt,mark options={line width=0.75pt},line join=round},
	  {color=android_red,line width=0.5pt,mark size=2pt,mark options={line width=0.75pt},line join=round},
	  {color=android_blue,line width=0.5pt,mark size=2pt,mark options={line width=0.75pt},line join=round},
	  {color=android_pink,line width=0.5pt,mark size=2pt,mark options={line width=0.75pt},line join=round},
	  {color=android_orange,line width=0.5pt,mark size=2pt,mark options={line width=0.75pt},line join=round},
	  {color=black,line width=0.5pt,mark size=2pt,mark options={line width=0.75pt},line join=round}}

\begin{document}

\title{Efficiently reducing transition curvature in heat-assisted magnetic recording with state-of-the-art write heads} 

\author{Christoph Vogler}
\email{christoph.vogler@tuwien.ac.at}
\affiliation{Institute of Solid State Physics, TU Wien, 1040 Vienna, Austria}
\affiliation{Institute of Analysis and Scientific Computing, TU Wien, 1040 Vienna, Austria}

\author{Claas Abert}
\author{Florian Bruckner}
\author{Dieter Suess}
\affiliation{Christian Doppler Laboratory for Advanced Magnetic Sensing and Materials, Institute for Solid State Physics, TU Wien, 1040 Vienna, Austria}

\begin{abstract}
The curvature of bit transitions on granular media is a serious problem for the read-back process. We address this fundamental issue and propose a possibility to efficiently reduce transition curvatures with state-of-the-art heat-assisted magnetic recording (HAMR) heads. We compare footprints of conventional with those of the proposed head design on different media, consisting of exchange coupled and single phase grains. Additionally, we investigate the impact of various recording parameters, like the full width at half maximum (FWHM) of the applied heat pulse and the coercivity gradient near the write temperature of the recording grains. The footprints are calculated with a coarse grained model, based on the Landau-Lifshitz-Bloch (LLB) equation. The presented simulations show a transition curvature reduction of up to 40\,\%, in the case of a medium with exchange coupled grains and a heat pulse with a FWHM of 40\,nm. We further give the reason for the straightening of the bit transitions, by means of basic considerations with regard to the effective recording time window (ERTW) of the write process. Besides the transition curvature reduction the proposed head design yields an improvement of the transition jitter in both down-track and off-track direction.
\end{abstract}

\maketitle 

\pgfplotsset{colormap/RdBu-9}

In recent years, heat-assisted magnetic recording~\cite{mayer_curiepoint_1958,mee_proposed_1967,guisinger_thermomagnetic_1971,kobayashi_thermomagnetic_1984,rottmayer_heat-assisted_2006} (HAMR) is about to become the future high density recording technology for which areal densities well beyond 1\,Tb/in$^2$ are achievable~\cite{wang_hamr_2013,wu_hamr_2013,weller_hamr_2014,vogler_areal_2016,rea_areal-density_2016,vogler_heat-assisted_2016}. Without heat-assist there does not exist any known possibility to straighten the curvature of written bit transition. Unfortunately, a pronounced curvature significantly deteriorates the signal-to-noise ratio of the recorded track \cite{zhu_correcting_2017}. For HAMR very recently a way was proposed to correct the curvature by means of a magnetic field, which increases from the track center to its edges~\cite{zhu_correcting_2017}. However, it is still an open question how to produce such a field shape on the length scale of several tens of nanometers. At least the development of a complex head design will be required.

In this work, we propose a method to correct transition curvatures in HAMR, without the need to change the state-of-the-art head design. The basic idea is to simply reverse the inductive write element or switch the roles of its pole tips.
\begin{figure}
   \includegraphics{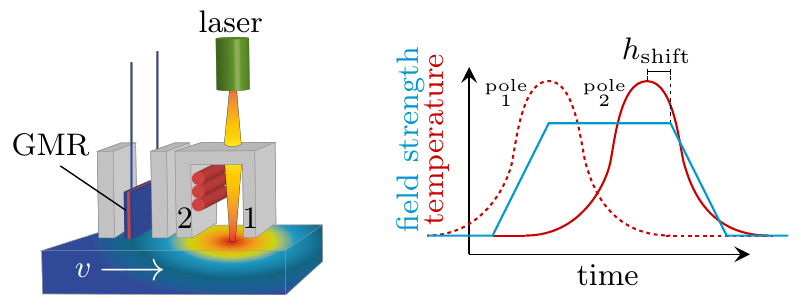}
      \caption{\small (color online) a) Schematic structure of a magnetoresistive HAMR head. b) Temporal evolution of temperature and magnetic field strength for a recording grain during HAMR. The situation, in which pole 1 is used as field source and the proposed situation with pole 2 as field source is compared.}
      \label{fig:sketch}  
\end{figure}
Figure~\ref{fig:sketch}a) schematically illustrates the structure of a magnetoresistive HAMR head, consisting of a GMR read sensor and an inductive write element with an additional laser, which provides the heat-assist. In conventional HAMR the granular medium is moved to the right (as indicated by the arrow). The temporal evolution of temperature and magnetic field at a recording grain is displayed in Fig.~\ref{fig:sketch}b). The grain gets first heated by the laser spot (dashed red curve) before the magnetic field (solid blue curve), produced by pole 1, reaches its maximum value. During cooling of the grain, which is the most important part in the write process, the field can be described in a good approximation as spatially homogeneous. If pole 2 is used as field source the situation changes. First, the field is applied and at its maximum magnitude the grain gets heated (solid red curve). Hence, during cooling of the grain the field already decreases. This allows to significantly reduce the transition curvature as we will show in the following. Note that the marked gap $h_{\mathrm{shift}}$ between the peak of the heat pulse and the start of the field decrease will turn out to be an important design parameter.

In our HAMR model we assume a continuous Gaussian shaped heat pulse with a write temperature of 720\,K, which moves with a velocity of $v=10$\,m/s over a granular medium. Depending on the off-track position each grain is subject to a heat pulse with different peak temperatures. Additionally, we assume that the pole tip produces an external magnetic field with a spatially trapezoidal shape in down-track direction, which encloses an angle of $6^{\circ}$ with the surface normal of the medium. The field gradient is 40\,mT/nm (or 400\,mT/ns in time) and the write frequency is 1\,GHz. This results in a magnetic write pulse of 1\,ns. If not otherwise specified a field strength of 0.8\,T and a full width at half maximum (FWHM) of the heat pulse of 40\,nm and $h_{\mathrm{shift}}=5$\,nm are assumed. 

\begin{figure}[!h]
\begin{adjustwidth}{-1cm}{-1cm}
\includegraphics{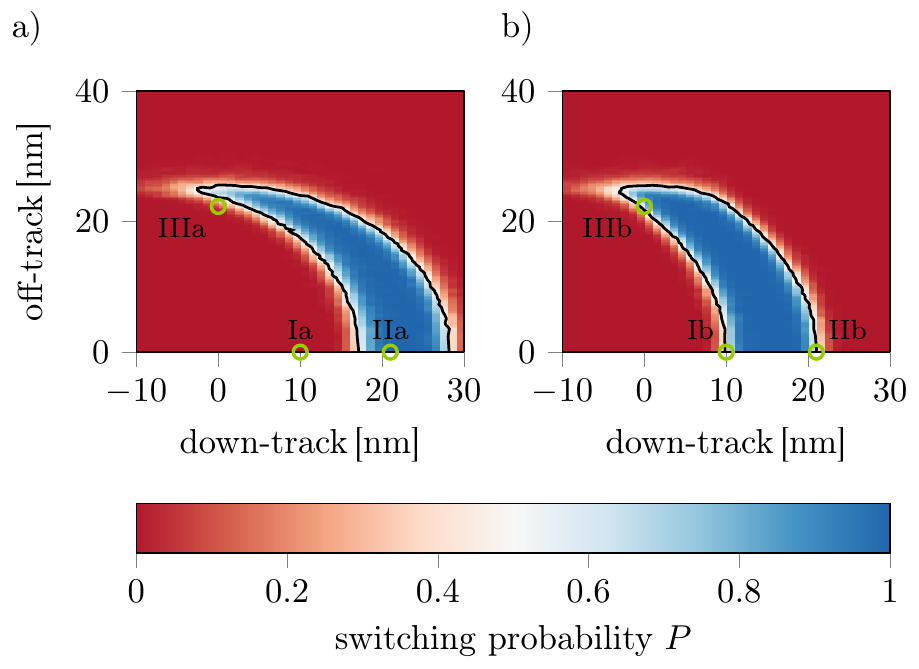}
  \caption{\small (color online) Footprint of a) a conventional (pole 1 in Fig.~\ref{fig:sketch}) b) the proposed HAMR write head (pole 2 in Fig.~\ref{fig:sketch}) with $\mathrm{FWHM}=40$\,nm and a write temperature of 720\,K on EC1 grains (see Table~\ref{tab:mat_match}). Further $v=10$\,m/s, $\mu_0 H_{\mathrm{ext}}=0.8$\,T and $h_{\mathrm{shift}}=5$\,nm in b) are assumed. The color code shows the switching probability obtained from 128 write trials simulated with a coarse grained LLB model (see Ref.\cite{volger_llb}). At a down-track position of 0\,nm the temporal evolution of the write pulse is centered around the Gaussian shaped heat pulse (as shown in Fig.~\ref{fig:track_edge}).}
  \label{fig:footprints}
\end{adjustwidth}
\end{figure}

Figure~\ref{fig:footprints} compares calculated footprints of the write head for a) the conventional case, in which pole 1 (see Fig.~\ref{fig:sketch}a) produces the field and b) the proposed case with pole 2 serving as field source. 
\begin{table}[h!]
  \centering
  \vspace{0.5cm}
  \small
  \begin{tabular*}{0.9\linewidth}{l @{\extracolsep{\fill}} ccc}
    \toprule
    \toprule
      & EC1 & EC2 & SP \\
    \midrule
    $K_{1,\mathrm{hard}}$\,[MJ/m$^3$] & $6.6$ &  $6.6$ & $6.6$\\
    $\mu_0 M_{\mathrm{S,hard}}$\,[T] & 1.43 & 1.43 & 1.43 \\
    $T_{\mathrm{C,hard}}$\,[K] & 644 & 537 & 537\\
    $K_{1,\mathrm{soft}}$\,[J/m$^3$] & 0.0 & 0.0 & -\\
    $\mu_0 M_{\mathrm{S,soft}}$\,[T] & 2.16 & 2.16 & - \\
    $T_{\mathrm{C,soft}}$\,[K] & 954 & 795 & - \\
    $\mathrm{d}H_{\mathrm{c}}/\mathrm{d}T|_{T \le T_{\mathrm{f,min}}}$\,[mT/K] & 4.8 & 5.6 & 24.2 \\
    \bottomrule
    \bottomrule
  \end{tabular*}
  \caption{\small Material parameters of two different exchange coupled grains EC1 and EC2 and a single phase grain SP. $K_1$ is the anisotropy constant, $M_{\mathrm{S}}$ the saturation magnetization, and $T_{\mathrm{C}}$ the Curie temperature of the used materials. $\mathrm{d}H_{\mathrm{c}}/\mathrm{d}T|_{T \le T_{\mathrm{f,min}}}$ denotes the average gradient of the coercive field for temperatures below $T_{\mathrm{f}}$, at which $H_{\mathrm{c}}$ becomes smaller than $H_{\mathrm{ext}}$.}
  \label{tab:mat_match}
\end{table}
The underlying medium is 10\,nm thick and consists of exchange coupled grains ($d=5$\,nm) with the material parameters EC1 in Table~\ref{tab:mat_match}. The HAMR head tries to reverse the magnetization of the grains from an initial down direction to the up direction. Figure~\ref{fig:footprints} shows the probability of switching $P$ of a grain as function of its down- and off-track position, after 128 write trials. Hence, it can be interpreted as the average footprint in a granular medium. 
The magnetization dynamics are computed with a coarse grained model based on the Landau-Lifshitz-Bloch (LLB) equation~\cite{garanin_thermal_2004,chubykalo-fesenko_dynamic_2006,atxitia_micromagnetic_2007,kazantseva_towards_2008,schieback_temperature_2009,bunce_laser-induced_2010,evans_stochastic_2012,mendil_resolving_2014}. Without going into too much detail, the model efficiently describes each material layer of a grain with a single magnetization vector. Nevertheless, the magnetization dynamics of an atomistically discretized model is correctly reproduced. Please refer to Ref.~\cite{volger_llb} for deeper insights. 

The simulated footprint of the proposed head with reversed pole tips in Fig.~\ref{fig:footprints}b shows a significant reduction of the transition curvature. At the track center the footprint is shifted by almost $-10$\,nm along the down-track direction, whereas at the track edge the bit transition almost remains at the same down-track position compared to the conventional head design of Fig.~\ref{fig:footprints}a. To quantify the curvature reduction we fit both trailing and leading edge of the footprints (line with a switching probability of 50\,\%) with a quadratic polynomial $cz^2+bz+a$~\cite{hashimoto_analysis_2007}. The second order coefficient $c$ is taken as measure for the curvature, yielding a large curvature reduction of almost 40\,\% for both edges of the footprint (see marked row in Table~\ref{tab:results}).
\begin{table}[h!]
  \centering
  \vspace{0.5cm}
  \small
  \begin{tabular*}{1.0\linewidth}{l c c c d{3.1} d{3.1} d{3.1} d{3.1}}
    \toprule
    \toprule
      & $\mu_0 H_{\mathrm{ext}}$ & FWHM & $h_{\mathrm{shift}}$ & \multicolumn{1}{c}{$c_{\mathrm{trailing}}$} & \multicolumn{1}{c}{$c_{\mathrm{leading}}$} & \multicolumn{1}{c}{$\sigma_{\mathrm{down}}$} & \multicolumn{1}{c}{$\sigma_{\mathrm{off}}$} \\
      & [T] & [nm] & [nm] & [\%] & [\%] & [\%] & [\%]\\
    \midrule
    EC1 & 0.8 & 20 & 0 & -\phantom{0}3.2 & -\phantom{0}4.9 & -17.2 & -10.2 \\
    EC1 & 0.8 & 20 & 5 & -23.3 & -23.1 & -22.1 & -\phantom{0}8.4 \\
    EC1 & 0.8 & 40 & 0 & -25.6 & -24.1 & -15.4 & -28.1 \\
    \rowcolor{android_dark_green!50}[1\tabcolsep]
    EC1 & 0.8 & 40 & 5 & -38.4 & -37.6 & -19.3 & -27.0 \\
    EC2 & 0.8 & 20 & 5 & -26.3 & -25.3 & -25.8 & -\phantom{0}1.6 \\
    EC2 & 0.8 & 40 & 5 & -39.0 & -40.4 & -36.8 & -33.6 \\
    SP & 1.2 & 20 & 5 & +\phantom{0}1.3 & -\phantom{0}4.2 & -28.7 & -\phantom{0}1.1 \\
    SP & 1.2 & 40 & 5 & -10.1 & -\phantom{0}6.8 & -49.8 & -43.7 \\
    \bottomrule
  \end{tabular*}
  \caption{\small Transition curvature reduction for the proposed head design with reversed pole tips compared to a conventional HAMR head. Various grains (see Table~\ref{tab:mat_match}) and design parameters are investigated. $\sigma_{\mathrm{down}}$ and $\sigma_{\mathrm{off}}$ denote the change in the transition jitter in down- and off-track direction, respectively.}
  \label{tab:results}
\end{table}

To get deeper insights into the mechanism of the presented reduction let us briefly recall the definition of the effective recording time window (ERTW) of Ref.~\cite{vogler_basic_2016}:
\begin{equation}
 \label{eq:ERTW}
 \text{ERTW}_\uparrow=\left[t(T_{\mathrm{C}}),t(T_{\mathrm{f}})\right] \cap \left [ t_{\uparrow,\mathrm{start}},t_{\uparrow,\mathrm{final}}\right].
\end{equation}
In HAMR recording grains get heated during the write process to reduce their coercive field. Above the freezing temperature $T_{\mathrm{f}}$ the coercivity decreases below a given write field strength. Hence, there only exists a small temperature range within which the magnetization of a grain can be reversed (between $T_{\mathrm{C}}$ and $T_{\mathrm{f}}$). The first term in Eq.~\ref{eq:ERTW} denotes precisely this time window. The ERTW for writing in up direction is then defined as intersection of the latter and the time period of the write pulse in up direction. The findings of Ref.~\cite{vogler_basic_2016} show that $\mathrm{ERTW}_{\uparrow}$ must be longer than a threshold value to obtain a switching probability of 100\,\%.

Based on this result we investigate the marked points in Fig.~\ref{fig:footprints} in more detail.
\begin{figure}
\begin{adjustwidth}{-1cm}{-1cm}
\includegraphics{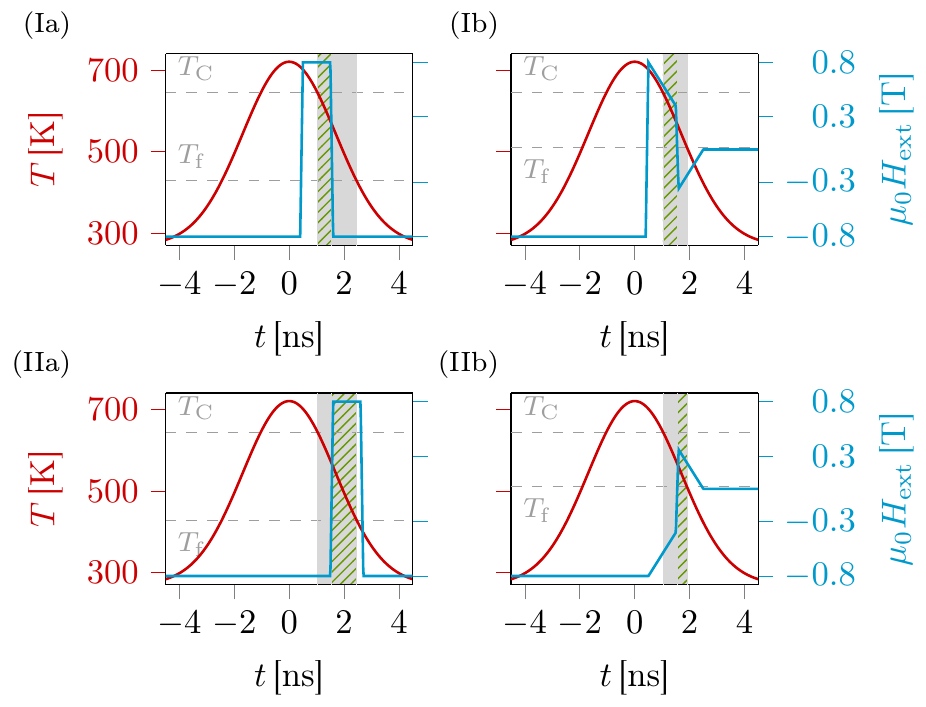}
    \caption{\small (color online) Temporal evolution of the applied heat pulse and the external field in the marked points of Fig.~\ref{fig:footprints}. The time span between $T_{\mathrm{C}}$ and $T_{\mathrm{f}}$ is shown with a gray solid area and $\mathrm{ERTW}_{\uparrow}$ corresponding to Eq.~\ref{eq:ERTW} is marked with a green striped area.}
  \label{fig:track_center}
\end{adjustwidth}
\end{figure}
Figure~\ref{fig:track_center} displays the temporal evolution of the heat pulse and the external field during the write process at the track center and compares the situation of the conventional and the proposed head design. At the track center the grains are subject to the full write temperature of 720\,K. 
According to Eq.~\ref{eq:ERTW} a grain's magnetization can only be aligned in up direction if the temperature is between $T_{\mathrm{C}}$ and $T_{\mathrm{f}}$ (gray solid area in Fig.~\ref{fig:track_center}) and the field points in up direction. The resulting $\mathrm{ERTW}_{\uparrow}$ is marked with a green striped area. Although a substantial $\mathrm{ERTW}_{\uparrow}$ can be seen in point Ia the switching probability is 0\,\% (see Fig.~\ref{fig:footprints}a). The reason is that the temperature is still above $T_{\mathrm{f}}$ after the field has reversed, and thus the magnetization can be overwritten in the initial down direction during $\mathrm{ERTW}_{\downarrow}$. The situation is different in the case of the proposed head design, because the field already decays with 400\,mT/ns during cooling. Due to the lower field $T_{\mathrm{f}}$ significantly increases (note, the lower the field the higher the required reduction of coercivity). Hence, the total recording time window narrows. As a result $\mathrm{ERTW}_{\uparrow}$ preserves its length ($T_{\mathrm{C}}$ remains the same) but $\mathrm{ERTW}_{\downarrow}$ gets smaller. Both ERTWs become equal in point Ib of Fig.~\ref{fig:track_center}, yielding a switching probability of about 50\,\%. In a nutshell, we demonstrated why the trailing transition at the track center shifts if we make use of the existing head field gradient based on the proposed head design.

The leading bit transition at the track center also shifts. As displayed in point IIa for a constant field strength a large $\mathrm{ERTW}_{\uparrow}$ is obtained, yielding a switching probability of 100\,\%. In contrast, in point IIb the field is already very small during the important time span. Hence, the increase of $T_{\mathrm{f}}$ again decreases the recording time window. In this point $\mathrm{ERTW}_{\uparrow}$ narrows below the threshold value, which is required for complete switching. As a consequence the leading bit transition is shifted compared to the conventional head design, again because of the field gradient.

\begin{figure*}
\includegraphics{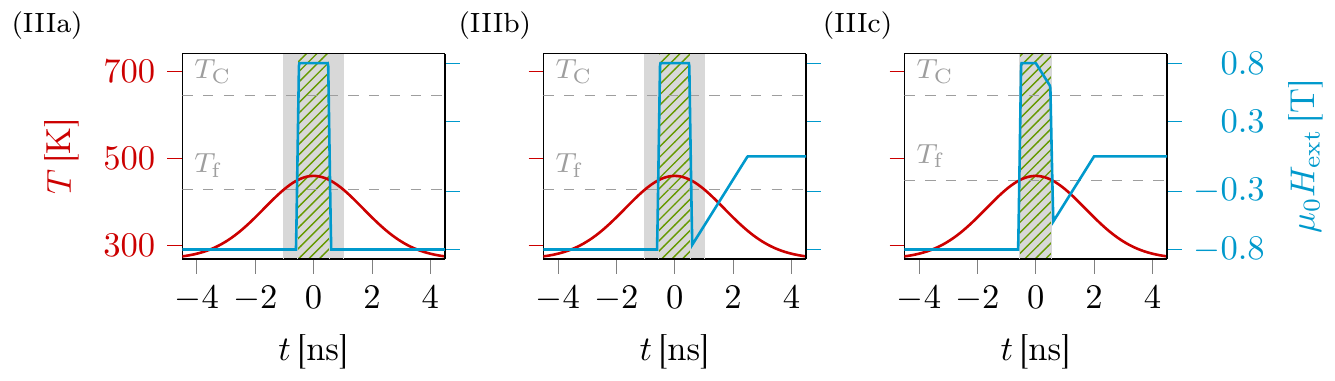}
    \caption{\small (color online) Same type of plot as shown in Fig.~\ref{fig:track_center} for the marked points in Fig.~\ref{fig:footprints}. Additionally the corresponding situation at a down-track position of 0\,nm and an off-track position of 22.5\,nm in the case of the proposed write head and $h_{\mathrm{shift}}=0$\,nm is shown in (IIIc).}
  \label{fig:track_edge}     
\end{figure*}

At the off-track edge of the footprint the peak temperature of the heat pulse is considerably lower than $T_{\mathrm{C}}$ as Fig.~\ref{fig:track_edge} shows.
The situation for the conventional (point IIIa) and the proposed (point IIIb) head design is almost identical. Due to the choice of $h_{\mathrm{shift}}=5$\,nm the field decay starts slightly after the write pulse. Hence, $T_{\mathrm{f}}$ does not change and both $\mathrm{ERTW}_{\uparrow}$ and $\mathrm{ERTW}_{\downarrow}$ for the two compared designs have almost the same length. As a consequence the bit transition does not shift at the off-track edge, which explains the overall reduction of the transition curvature for the proposed head design. The importance of the design parameter $h_{\mathrm{shift}}$ can be seen in point IIIc of Fig.~\ref{fig:track_edge}. Here, $h_{\mathrm{shift}}=0$\,nm is chosen, which means that the field starts to decay at the peak of the heat pulse (compare Fig.~\ref{fig:sketch}b). For this reason $T_{\mathrm{f}}$ slightly increases which results in a vanishing $\mathrm{ERTW}_{\downarrow}$, and thus no overwriting occurs. This indicates that the point is already located in the middle of the footprint with a switching probability of 100\,\%, which means that the off-track edge shifts to the left, reducing the overall curvature reduction. Table~\ref{tab:results} confirms these considerations, since the curvature reduction drastically decreases for the same recording parameters but $h_{\mathrm{shift}}=0$\,nm. Depending on the detailed field shape and the velocity $v$, $h_{\mathrm{shift}}$ can be controlled by the distance between the heat spot center and the pole tip.

To further examine the transition curvature reduction of the proposed write head footprints for a heat pulse with $\mathrm{FWHM}=20$\,nm are computed. The reduction decreases for decreasing FWHM as Table~\ref{tab:results} points out. The reason is that the absolute transition curvature lowers if the heat spot narrows independent from the head design. Hence, the transition shift along the center of the track is less pronounced. For $h_{\mathrm{shift}}=5$\,nm the reduction of the transition curvature almost halves and for $h_{\mathrm{shift}}=0$\,nm the shift of the footprint at the track center is almost identical to that at the off-track edge, yielding a marginal curvature reduction. This fact again points out the importance of the design parameter $h_{\mathrm{shift}}$. 

In the case of exchange coupled grains EC2 with lower Curie temperatures in both the soft and the hard magnetic part, the curvature reductions are very similar to those of EC1 grains. Obviously, the Curie temperature does not influence the mechanism, if the gradients of the coercive field $\mathrm{d}H_{\mathrm{c}}/\mathrm{d}T$ around the freezing temperature are similar. To check how the transition curvature reduction changes if grains with a larger $\mathrm{d}H_{\mathrm{c}}/\mathrm{d}T$ gradient are used, we analyze a hard magnetic single phase grain SP. Here, $\mathrm{d}H_{\mathrm{c}}/\mathrm{d}T$ is about five times larger than for the exchange coupled grains EC1 and EC2. For the SP grains a higher external magnetic field of 1.2\,T is used, otherwise the calculated footprints show DC noise. However, the used field gradient remains unchanged with 40mT/nm. Table~\ref{tab:results} points out that the curvature reduction significantly decreases for $\mathrm{FWHM}=40$\,nm. The reason is that the actual freezing temperature is much higher for the SP grain compared to the EC2 grain with the same Curie temperature in the hard magnetic part. This is a direct consequence of the large $\mathrm{d}H_{\mathrm{c}}/\mathrm{d}T$ gradient. Hence, the available ERTW is much smaller, which weakens the effect described in Fig.~\ref{fig:track_center}, and thus limits the shift of the bit transitions at the track center. In the case of $\mathrm{FWHM}=20$\,nm the curvature reduction even vanishes. This is not surprising because, as mentioned above, for a narrowing heat profile the absolute transition curvature is decreasing independently from the head design.

Besides the transition curvature reduction a substantial decrease of the transition jitter in down-track direction is obtained as displayed in Table~\ref{tab:results}. Since, for the proposed head design thermal and field gradient actually add up, the effective head field gradient
\begin{equation}
\label{eq:write_field}
 \frac{\mathrm{d}H_{\mathrm{eff}}}{\mathrm{d}x}=\frac{\mathrm{d}|H_{\mathrm{ext}}|}{\mathrm{d}x}+\frac{\mathrm{d}H_{\mathrm{c}}}{\mathrm{d}T} \frac{\mathrm{d}T}{\mathrm{d}x}
\end{equation}
increases. Hence, the improvement of the transition jitter depends on both the FWHM and the $\mathrm{d}H_{\mathrm{c}}/\mathrm{d}T$ gradient, which is in good agreement with the values of Table~\ref{tab:results}. Additionally, due to the straighting of the footprints a reduction of the off-track transition jitter results as a side effect.

In summary, we demonstrated an efficient way to reduce the curvature of bit transitions on granular media with state-of-the-art HAMR heads. The idea relies on the reversal of the inductive write element. As a consequence this design produces a decaying magnetic field strength during the effective recording time window ERTW between $T_{\mathrm{C}}$ and the freezing temperature $T_{\mathrm{f}}$ at which the coercive field equals the given magnetic write field. The field gradient shifts bit transitions at the center of the track, but not at the edge of the track, which straightens the transition. This works best for large heat spots, because the decaying magnetic field confines the bit. For exchange coupled grains with a moderate gradient of the coercivity near $T_{\mathrm{f}}$ and a FWHM of 40\,nm of the heat pulse, we obtained a curvature reduction of about 40\,\%. The effect lessens for decreasing 
FWHM and increasing coercivity gradient. Hence, it is less important for ultrahigh density devices with areal densities $>5$\,Tb/in$^2$, in which the transition curvature will naturally decrease due to the need of materials with a high coercivity gradient and heat spots with a low FWHM. However, the proposed head design can significantly increase the signal-to-noise ratio of current HAMR devices, suffering from the curvature of bit transitions. As a side effect the proposed head design has a positive impact on the transition jitter in both down-track and off-track direction.

The authors would like to thank the Vienna Science and Technology Fund (WWTF) under grant No. MA14-044, the Advanced Storage Technology Consortium (ASTC), and the Austrian Science Fund (FWF) under grant No. I2214-N20 for financial support. The computational results presented have been achieved using the Vienna Scientific Cluster (VSC).
\bibliography{/home/christoph/Dropbox/HAMR}

\end{document}